\documentclass[letterpaper, 10 pt, conference]{ieeeconf}  
\usepackage{cite}
\usepackage{float}
\usepackage{multirow, tabularx}
\usepackage{graphicx}
\usepackage{soul,color}
\graphicspath{ {./images/} }
\IEEEoverridecommandlockouts                              

\title{\LARGE \bf
Measuring Cognitive Workload Using Multimodal Sensors

}

\author{Niraj Hirachan$^{1}$, Anita Mathews$^{1}$, Julio Romero$^{1}$, and Raul Fernandez Rojas$^{1*}$
\thanks{$^{1}$Human-Centred Technology Research Centre, Faculty of Science and Technology, University of Canberra, 11 Kirinari St, Bruce ACT, Australia
        {\tt\small *corresponding author: raul.fernandezrojas@canberra.edu.au}}%
}

\begin{document}

\maketitle
\thispagestyle{empty}
\pagestyle{empty}

\begin{abstract}
This study aims to identify a set of indicators to estimate cognitive workload using a multimodal sensing approach and machine learning. A set of three cognitive tests were conducted to induce cognitive workload in twelve participants at two levels of task difficulty (Easy and Hard). Four sensors were used to measure the participants' physiological change, including, Electrocardiogram (ECG), electrodermal activity (EDA), respiration (RESP), and blood oxygen saturation (SpO2). To understand the perceived cognitive workload, NASA-TLX was used after each test and analysed using Chi-Square test. Three well-know classifiers (LDA, SVM, and DT) were trained and tested independently using the physiological data. The statistical analysis showed that participants' perceived cognitive workload was significantly different ($p<0.001$) between the tests, which demonstrated the validity of the experimental conditions to induce different cognitive levels. Classification results showed that a fusion of ECG and EDA presented good discriminating power (acc~=~0.74) for cognitive workload detection. This study provides preliminary results in the identification of a possible set of indicators of cognitive workload. Future work needs to be carried out to validate the indicators using more realistic scenarios and with a larger population.\newline
\end{abstract}

\section{INTRODUCTION} 
Humans have a limited amount of cognitive resources. At any instance, we can process only a certain amount of information and maintaining a healthy cognitive function is a challenge. For instance, continuous high cognitive workload (i.e., overload) for extended periods have a negative impact on performance, and sub-optimal decisions, human errors, or accidents might occur~\cite{fernandez2020electroencephalographic}. Similarly, low levels of cognitive load (i.e., underload) can affect performance, due to lack of concentration, boredom, or lost of motivation. Thus, it is important to evaluate the cognitive load imposed by different tasks in objective terms and in real time. 
This measurement would enable us to create assistive systems, that can assist the user in making optimal decisions when faced in overload or underload conditions, and also maintain a healthy psychological well-being in the long run. 


In the literature, two main methods to measure cognitive load have been proposed: subjective and objective measures~\cite{fernandez2020electroencephalographic}. Subjective measures are based on the self-reporting scales, questionnaires, or interviews. A popular metric is the NASA Task Load Index (NASA-TLX) questionnaire, since it is a well-established and reliable tool to measure workload \cite{galy2018measuring}. However, a disadvantage of this technique is that it lacks the measurement of the cognitive load objectively in real time as the questionnaire is completed at the end of the task. On the other hand, objective physiological measures are based on collecting physiological signals from sensors. This approach is based on the principle that the cognitive load is reflected in physiological signals controlled by the autonomic nervous system ~\cite{charles2019measuring}. 


Applications of physiological sensors to measure cognitive load are found in the literature. For instance, Nourbakhsh et al.~\cite{nourbakhsh2013gsr}, used EDA and eye blinks to assess cognitive workload during an arithmetic task. 
In another example, Tsunoda et al.~\cite{tsunoda2015estimating}, employed heart rate variability (HRV) to examine cognitive load in the advanced trail making test. 
Heeman et al.~\cite{heeman2013estimating}, used pupillary diameter for the estimation of cognitive load during a dialog task. While research has been done in measuring the cognitive load using physiological sensors, these studies have been limited to a certain type of activity. However, there is the need to investigate multiple physiological correlates in multiple cognitive tasks.





In this study, we designed an experiment where we used four different physiological indicators to measure cognitive workload in twelve participants while performing multiple tasks. The three cognitive tasks are Raven test~\cite{carpenter1990one}, Numerical test, and a Video Game; all test under two type of difficulty level ($Easy$ and $Hard$). 
Given the importance of cognitive workload in human performance, mental fatigue, and detrimental effect to mental health, measuring cognitive workload objectively is imperative in managing mental resources and maintaining overall well-being.



\section{METHODOLOGY}

\subsection{Participants} 
Twelve participants (7M/5F) took part in the experiment. Their age ranged from 20 to 40 year old (mean age 25 $\pm$ 5.5 std). Written informed consent was obtained before the experiment and it was informed that no personal information would be recorded. The experimental procedures involving human subjects described in this paper were approved by the Institutional Review Board. Prior to the start of the experiment, the protocol was clearly explained.

\subsection{Experimental Protocol}
\begin{figure}[h]

\centering
\includegraphics[width=6cm]{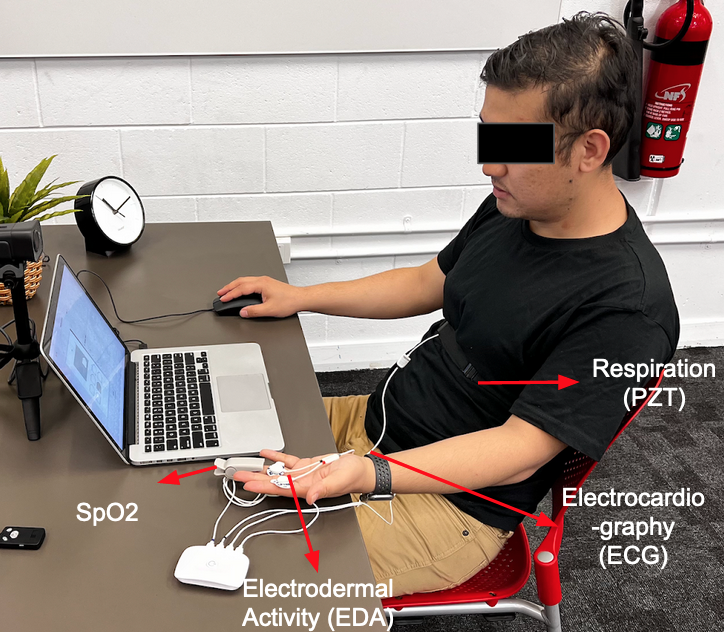}
\caption{Experiment setup with subject wearing the physiological sensors}
\label{fig:experiment_setup} 
\end{figure}

All experiments were conducted at the Human-Machine Interface Laboratory at University of Canberra, Australia. During the experiment, the participants were seated on a fixed chair in front of a computer screen placed on a desk. Four sensors (Biosignal plux, Lisbon, Portugal) were used to record physiological data: Electrocardiography (ECG), Electrodermal Activity (EDA), Respiration (RESP), and Oxigen Saturation (SpO2). Figure \ref{fig:experiment_setup} presents an example of the experimental setup. 



The experiment consists of 3 different cognitive activities (Game, Numerical Test and Raven's Test) to induce cognitive load. Each activity is divided into two difficulty levels, \textit{Easy} and \textit{Hard}; in total there were 6 activities (3 Tests $\times$ 2 levels). In Game, the participants played a bouncing ball video game where they have to navigate through a maze by clicking a mouse button; the difficulty level was obtained by increasing the speed of the game. Raven's test consisted of 10 visual items depicting a matrix of colored geometric shapes arranged in a 1x6 layout, each matrix contained one empty cell and options for the participant to select from to complete the matrix; two difficulty levels were taken from \cite{raven2003raven}. The numerical test included 10 multi-choices for the participants to solve mentally; the \textit{Easy} level included operations with 2 variables, while the \textit{Hard} level included 4 variables. 

The whole experiment was divided into two parts following a randomized orthogonal experimental design. At the start of the experiment one minute was given for baseline recording, after that, the experiment was presented to the participants with each activity lasting two minutes. After completing each activity, a computer-based NASA-TLX assessment was administered to the participants. The NASA-TLX is used to measure the subjective mental load of the participants (ground truth of experiment)\cite{fernandez2020electroencephalographic}. The NASA-TLX measures 6 different aspects:  Mental Demand, Physical Demand, Temporal Demand, Own Performance, Effort and Frustration level.

\subsection{Data Analysis}

\subsubsection{Validation of Experimental Conditions}   
In order to validate the design of this experiment and the experimental conditions, the response of the NASA-TLX questionnaire was analysed. Thus, the null hypothesis is that the perceived workload is not affected during the experimental task. A Chi-Square test of contingencies (with $\alpha = 0.05$) was used to assess if the NASA-TLX items were related to the different experimental conditions (\textit{Easy} and \textit{Hard} levels) across all the participants. 
The assumptions for independence and minimum expected frequencies for the statistical test were met. 
Post-hoc tests were undertaken using a Bonferroni Correction for pairwise comparisons between tests. 

\subsubsection{Pre-processing}  
All sensors collected physiological data with a sampling rate of $100Hz$. The data was processed using 10s windows, as this window size was the optimal after testing several window sizes (5s-50s). Then, each physiological signal was treated separately to remove noise. For ECG signal, a band-pass filter ($0.5-40Hz$) was used to remove high-frequency oscillations and powerline interference. Similarly, the respiration data was filtered with a band-pass filter ($0.05-0.5Hz$). For the EDA data, a low-pass filter ($5Hz$) was applied to remove line noise.




\subsubsection{Feature Extraction} 
Physiological features that potentially correlate with cognitive workload were extracted from the four sensors. Statistical features (n=10) were obtained from EDA and SPO2 sensors, including: Mean, Standard Deviation, Amplitude, Trough, Range, Skew, Kurtosis, first and third quartiles (Q1, Q3), and interquartile range (IQR). From ECG signals, morphological features (n=10) were calculated: R-peaks, beats per minute, inter-beat interval (IBI), standard deviation of RR intervals (stdI), standard deviation of successive differences (stdD), root mean square of successive differences, proportion of successive differences above 20ms (PNN20), proportion of successive differences above 50ms (PNN50), median absolute deviation of RR intervals (MAD), and breathing rate. From the respiration signal (n=11), inhalation, exhalation, range estimation top and bottom percentile of peaks, trough respiratory values, offset level, slope, breath-to-breath interval (BBI), standard deviation, Q1, Q3, and IQR were calculated. In total we extracted 41 features.

\begin{figure*}[ht]
\centering
\includegraphics[width=17cm]{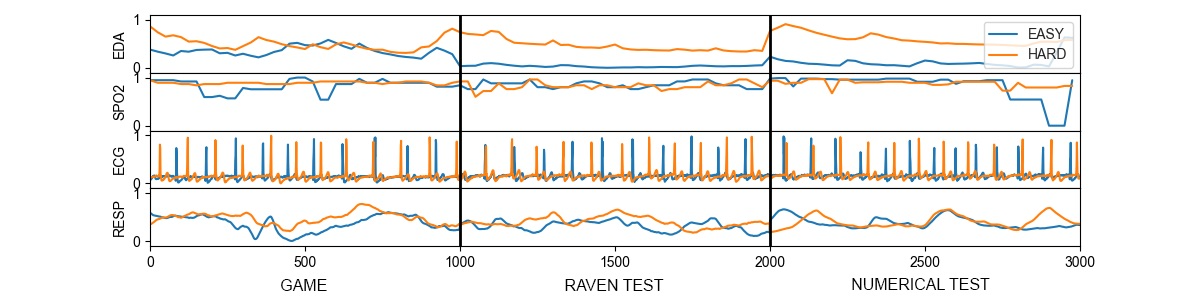}
\caption{An example of the physiological data captured by the EDA, SpO2, ECG, and Respiration sensors during each of the experimental conditions. }
\label{fig:sensors}
\end{figure*} 

\subsubsection{Feature Selection}
For all the sensor data, appropriate feature extraction was conducted to extract the best features and to build a more accurate model. We chose Joint Mutual Information (JMI) for feature selection, 
as it presents a good trade-off in terms of accuracy, stability, and flexibility than other ranking methods \cite{rojas2019machine}. An early fusion approach was followed to concatenate all computed features before the classification task.

\subsubsection{Classification}
The objective of the classification problem was to identify the difficulty level ($Easy$ vs $Hard$) from the physiological data. Three kinds of classification methods are compared to establish more appropriate recognition models, these are: Linear Discriminant Analysis (LDA), Support Vector Machine (with radial basis kernel), and Decision Trees. For all the classifiers, the data was split into 70\% training and remaining 30\% testing. A 10-Folds cross-validation was performed during the training process. 


\section{RESULTS}
Figure \ref{fig:sensors} presents an example of the recorded data from all four sensors in each experimental condition.


\subsection{Validation of Experimental Conditions}
The experimental assumption is that in \textit{Hard} experimental conditions, the participant's perceived workload will be significant different than in \textit{Easy} conditions. Following a Chi-Square test, the Game-$\chi^2$ test was non-significant $\chi^2(5,731)=1.363$, $p=0.928$. On the other hand, the Raven's-$\chi^2$ test was statistically significant $\chi^2(5,710)=13.253$, $p=0.021$. Similarly, the Numerical-$\chi^2$ test was also statistically significant $\chi^2(5,827)=11.065$, $p=0.031$. 
Pairwise comparisons reported that the differences between \textit{Easy} and \textit{Hard} tasks were significant for Raven's Frustration, $\chi^2(1,710)=46.24$, $p<0.000$; Mental Demand, $\chi^2(1,710)=51.84$, $p<0.000$; Performance, $\chi^2(1,710)=342.25$, $p<0.000$; and Temporal Demand, $\chi^2(1,710)=16.81$, $p<0.000$. Similarly, significant differences in Numerical test sub-scales were found in Performance $\chi^2(1,827)=18.40$, $p<0.000$ and Mental Demand, $\chi^2(1,827)=15.10$, $p=0.0024$. 

   \begin{figure*}[t]
      \centering
      \includegraphics[width=17cm]{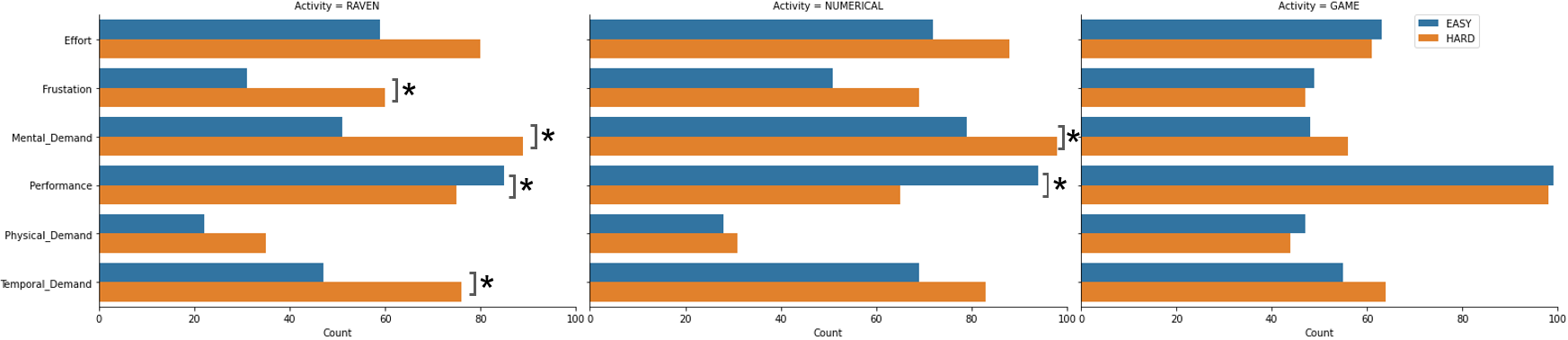}
      \caption{Pairwise Comparisons Among Tasks and NASA-TLX sub-scales. * $p<0.001$}
      \label{barChart}
   \end{figure*}

\subsection{Evaluation of Physiological Indicators}

\subsubsection{Baseline Results} 
The results on the test set using LDA, SVM and DT are presented in Table \ref{baseline_table}. The classification models were applied with data from each sensor individually and then with all sensors combined, the number of features used in each model appears in parenthesis. When analysed separately, Respiration and SpO2 sensor data obtained the lowest accuracy using the LDA (acc = 0.60) and DT (acc = 0.56), respectively. On the other hand, the ECG and EDA sensor data obtained favorable results (acc = 0.68) using DT. Overall, the best results were obtained using data from all four sensors with DT (acc = 0.70). 

\begin{table}[h]
    \centering
     \caption{Baseline results, number of features appear in parentheses. }
    \begin{tabular}{c c c c c  c} 
    \hline
  	&	ECG (10)	&	RESP (11)	&	Sp02 (10)	&	EDA (10)	&	All (41)	\\
  	\hline
LDA	&	0.63	&	\textbf{0.60}	&	0.53	&	0.61	&	0.67	\\
SVM	&	0.66	&	\textbf{0.60}	&	0.51	&	0.62	&	0.66	\\
DT	&	\textbf{0.68}	&	0.57	&	0.56	&	\textbf{0.68}	&	\textbf{0.70}	\\
\hline
    \end{tabular} 
    \label{baseline_table}
\end{table}
 
\subsubsection{Feature Selection}
\begin{figure}[h]
      \centering
      \includegraphics[width=7cm]{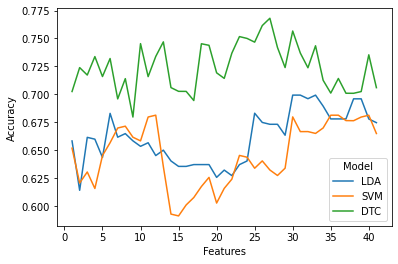}
      \caption{ Classification results using ranked features.}
      \label{fig:result_jmi}
   \end{figure}

After using JMI to find feature significance, all features were ranked and used to train and test the classifiers. The identification of the best performing model was systematically tested with a different number of features based on their feature importance (ranking). Figure \ref{fig:result_jmi} presents the results of this systematic search. It is clear that DT obtained the best results (acc = 0.77) with 27 features. Lower results were obtained with the LDA (acc = 0.70) and the SVM (acc = 0.68) classifiers using 30 and 12 features, respectively. However, by using the top-10 most important features, the DT classifier obtained acceptable results (acc = 0.74), which represents a clear improvement from the baseline results (acc = 0.70) using the 41 features from all sensors. These top-10 features are (in order of importance): $EDA\_mean$, $EDA\_Q3$, $EDA\_min$, $ECG_PNN50$, $ECG\_PNN20$, $EDA\_max$, $ECG\_IBI$, $ECG_MAD$, $ECG\_stdI$.


       
	

\section{Discussions}
The subjective workload assessment using NASA-TLX responses was evaluated to determine if the experimental conditions induced different levels of cognitive load. Based on the statistical analysis, the null hypothesis was rejected; thus, it was found that in \textit{Hard} experimental conditions, the participants' perceived workload was significant different than in \textit{Easy} conditions. Pairwise comparison showed that not all experiments induced different levels of cognitive load. The statistical significance for Raven’s and Numerical tests indicated that both were able to detect differences between metrics of the NASA-TLX; however, the Game test did not contributed on differentiating Hard and Easy levels of cognitive load. It can be argued that the reason there are not statistical differences in the items of the Game test is because the majority of the participants were university students, which are familiar with video games. This test was considered less difficult than expected due to the experience of the users. Hence, in our future work we will increase the difficulty level to make the game more challenging for the participants.


The statistical tests also showed significant differences in the NASA-TLX sub-scales. For instance, most of the sub-scales for the Raven's test were significantly different, with the perceived Frustration, Mental Demand, Performance, and the Temporal Demand sub-scales higher in the $Hard$ condition than in the $Easy$ condition. Similarly, the Numerical test exhibited significant differences in the perceived Mental Demand and Performance sub-scales. In these two tests, the perceived Mental Demand was the highest sub-scale during the \textit{Hard} (high difficulty) condition; while perceived reported Performance was higher in the \textit{Easy} (low difficulty) condition. These results are in line with similar studies where NASA-TLX has been employed to assess workload variations in different conditions \cite{lowndes2020nasa}.

Physiological data were used to investigate the automated identification of cognitive workload. Using data from each sensor separately, both the ECG and EDA outperformed (acc = 0.68) the RESP (acc = 0.60) and SpO2 (acc = 0.56) sensors. The performance of the ECG and EDA sensor data was further confirmed after the feature selection process, in which the top-10 features were composed only with features from ECG and EDA data (acc = 0.74). The obtained results are comparable with published results in the literature. For instance, Ding et el., \cite{ding2020measurement}, found that a fusion of ECG and EDA reached satisfactory results (acc = $0.58\%$) for the classification of mental workload using neural networks. In another similar study \cite{xu2014cluster}, a combination of ECG, GSR, SpO2, electroencephalography, and electromyography were successfully used to discriminate different cognitive tasks using SVM (acc = $0.73$). These results show that ECG and EDA are possible indicators of cognitive workload.   

Finally, the contributions of this study can be summarised as follows: 1) it offers an exploratory study that aims to compare different physiological metrics for the objective assessment of cognitive workload, and 2) it presents ten physiological indicators from both ECG and EDA as potential indicators for the objective assessment of cognitive workload.






\section{CONCLUSIONS}
This study aimed to determine a set of indicators to estimate cognitive workload using physiological sensors and machine learning. Ten features were identified as possible indicators of cognitive workload. Fusion ECG and EDA data were more informative than RESP and SpO2 data to discriminate between cognitive workload levels; fusion of multiple sensors usually improves cognitive workload assessment~\cite{debie2019multimodal,webber2021human}. In our future work, these indicators will be tested and validated using other cognitive tasks, more realistic scenarios, and with a larger population.

\addtolength{\textheight}{-12cm}   



\bibliographystyle{IEEEtran}
\bibliography{biblography}

\begin{thebibliography}{10}
\providecommand{\url}[1]{#1}
\csname url@samestyle\endcsname
\providecommand{\newblock}{\relax}
\providecommand{\bibinfo}[2]{#2}
\providecommand{\BIBentrySTDinterwordspacing}{\spaceskip=0pt\relax}
\providecommand{\BIBentryALTinterwordstretchfactor}{4}
\providecommand{\BIBentryALTinterwordspacing}{\spaceskip=\fontdimen2\font plus
\BIBentryALTinterwordstretchfactor\fontdimen3\font minus
  \fontdimen4\font\relax}
\providecommand{\BIBforeignlanguage}[2]{{%
\expandafter\ifx\csname l@#1\endcsname\relax
\typeout{** WARNING: IEEEtran.bst: No hyphenation pattern has been}%
\typeout{** loaded for the language `#1'. Using the pattern for}%
\typeout{** the default language instead.}%
\else
\language=\csname l@#1\endcsname
\fi
#2}}
\providecommand{\BIBdecl}{\relax}
\BIBdecl

\bibitem{fernandez2020electroencephalographic}
R.~Fernandez~Rojas, E.~Debie, J.~Fidock, M.~Barlow, K.~Kasmarik, S.~Anavatti,
  M.~Garratt, and H.~Abbass, ``Electroencephalographic workload indicators
  during teleoperation of an unmanned aerial vehicle shepherding a swarm of
  unmanned ground vehicles in contested environments,'' \emph{Frontiers in
  neuroscience}, vol.~14, p.~40, 2020.

\bibitem{galy2018measuring}
E.~Galy, J.~Paxion, and C.~Berthelon, ``Measuring mental workload with the
  nasa-tlx needs to examine each dimension rather than relying on the global
  score: an example with driving,'' \emph{Ergonomics}, vol.~61, no.~4, pp.
  517--527, 2018.

\bibitem{charles2019measuring}
R.~L. Charles and J.~Nixon, ``Measuring mental workload using physiological
  measures: A systematic review,'' \emph{Applied ergonomics}, vol.~74, pp.
  221--232, 2019.

\bibitem{nourbakhsh2013gsr}
N.~Nourbakhsh, Y.~Wang, and F.~Chen, ``Gsr and blink features for cognitive
  load classification,'' in \emph{IFIP conference on human-computer
  interaction}.\hskip 1em plus 0.5em minus 0.4em\relax Springer, 2013, pp.
  159--166.

\bibitem{tsunoda2015estimating}
K.~Tsunoda, A.~Chiba, H.~Chigira, T.~Ura, and O.~Mizunq, ``Estimating changes
  in a cognitive performance using heart rate variability,'' in \emph{2015 IEEE
  15th International Conference on Bioinformatics and Bioengineering
  (BIBE)}.\hskip 1em plus 0.5em minus 0.4em\relax IEEE, 2015, pp. 1--6.

\bibitem{heeman2013estimating}
P.~A. Heeman, T.~Meshorer, A.~L. Kun, O.~Palinko, and Z.~Medenica, ``Estimating
  cognitive load using pupil diameter during a spoken dialogue task,'' in
  \emph{Proceedings of the 5th International Conference on Automotive User
  Interfaces and Interactive Vehicular Applications}, 2013, pp. 242--245.

\bibitem{carpenter1990one}
P.~A. Carpenter, M.~A. Just, and P.~Shell, ``What one intelligence test
  measures: a theoretical account of the processing in the raven progressive
  matrices test.'' \emph{Psychological review}, vol.~97, no.~3, p. 404, 1990.

\bibitem{raven2003raven}
J.~Raven \emph{et~al.}, ``Raven progressive matrices,'' in \emph{Handbook of
  nonverbal assessment}.\hskip 1em plus 0.5em minus 0.4em\relax Springer, 2003,
  pp. 223--237.

\bibitem{rojas2019machine}
R.~F. Rojas, X.~Huang, and K.-L. Ou, ``A machine learning approach for the
  identification of a biomarker of human pain using fnirs,'' \emph{Scientific
  reports}, vol.~9, no.~1, pp. 1--12, 2019.

\bibitem{lowndes2020nasa}
B.~R. Lowndes, K.~L. Forsyth, R.~C. Blocker, P.~G. Dean, M.~J. Truty, S.~F.
  Heller, S.~Blackmon, M.~S. Hallbeck, and H.~Nelson, ``Nasa-tlx assessment of
  surgeon workload variation across specialties,'' \emph{Annals of surgery},
  vol. 271, no.~4, pp. 686--692, 2020.

\bibitem{ding2020measurement}
Y.~Ding, Y.~Cao, V.~G. Duffy, Y.~Wang, and X.~Zhang, ``Measurement and
  identification of mental workload during simulated computer tasks with
  multimodal methods and machine learning,'' \emph{Ergonomics}, vol.~63, no.~7,
  pp. 896--908, 2020.

\bibitem{xu2014cluster}
Q.~Xu, T.~L. Nwe, and C.~Guan, ``Cluster-based analysis for personalized stress
  evaluation using physiological signals,'' \emph{IEEE journal of biomedical
  and health informatics}, vol.~19, no.~1, pp. 275--281, 2014.

\bibitem{debie2019multimodal}
E.~Debie, R.~F. Rojas, J.~Fidock, M.~Barlow, K.~Kasmarik, S.~Anavatti,
  M.~Garratt, and H.~A. Abbass, ``Multimodal fusion for objective assessment of
  cognitive workload: a review,'' \emph{IEEE transactions on cybernetics},
  vol.~51, no.~3, pp. 1542--1555, 2019.

\bibitem{webber2021human}
M.~Webber and R.~F. Rojas, ``Human activity recognition with accelerometer and
  gyroscope: a data fusion approach,'' \emph{IEEE Sensors Journal}, 2021.

\end{thebibliography}

\end{document}